\documentclass{article} 
\newtheorem{theorem}{Theorem}[section]

\newcommand{\RR}{\mathbf}

\usepackage{xspace}
\usepackage{amsfonts} 
\newcommand{\func}[1]{\ensuremath{\textsc{#1}}\xspace}

\newcommand{\sOB}{\ensuremath{\widetilde{\mathcal{O}}_B}}
\newcommand{\sO}{\ensuremath{\widetilde{\mathcal{O}}}}
\newcommand{\OB}{\ensuremath{{\mathcal{O}}_B}} 

\begin{document}

\noindent

{\bf Computing Handbook (Third edition) 

\bigskip 

\bigskip 

Volume I: Computer Science
           and Software Engineering 

\bigskip 

\bigskip 
 
Allen Tucker, Teo Gonzales and Jorge L. Diaz-Herrera, editors

\bigskip 

\bigskip

Taylor and Francis Group Publishers, 2014}

\clearpage
\centerline{
{\bf 
CHAPTER 10: ALGEBRAIC 
ALGORITHMS}}\footnote{
This material is based on work supported in part
by the European Union through
Marie-Curie Initial Training Network ``SAGA'' (ShApes, Geometry, Algebra),
with FP7-PEOPLE contract PITN-GA-2008-214584 (first author),
by NSF Grant CCF-1116736 and 
PSC CUNY Awards 63153--0041 and 64512--0042 (second author), 
by  
the Danish Agency for Science, Technology and Innovation
(postdoctoral
grant), 
Danish NRF and NSF 
of China (grant 61061130540),
CFEM, 
and the Danish
Strategic Research Council (third author).
Sections 3.5, 5, and ``Further information" 
have been written jointly by all authors,
Section 4 has been contributed essentially by the first author, 
the other sections  
by the second author.}

\hspace*{0.2cm}

\smallskip

\hspace*{0.2cm}

\noindent
Ioannis Z.\ Emiris  \\
Department of Informatics and Telecommunications,\\
University of Athens, Athens 15784, Greece.
{\tt emiris@di.uoa.gr}
\hspace*{0.02cm}

\smallskip

\noindent
Victor Y.\ Pan\\
Mathematics and Computer Science Department, Lehman College,\\
City University of New York, Bronx, NY 10468, USA.
{\tt vpan@lehman.cuny.edu}.\\
{\tt http://comet.lehman.cuny.edu/vpan/} 
\hspace*{0.02cm}

\smallskip

\noindent
Elias P.\ Tsigaridas  \\
project PolSys, Laboratoire d' Informatique de Paris 6
Université Pierre and Marie Curie and INRIA Paris-Rocquencourt, France
{\tt elias@polsys.lip6.fr}
\hspace*{0.02cm}



\section{Introduction}

Algebraic algorithms deal with numbers, vectors,  
matrices, polynomials, formal power series,
exponential and differential polynomials, rational functions,
algebraic sets, curves and surfaces.
In this vast area, manipulation
with matrices and polynomials   
is fundamental for modern computations in Sciences and Engineering.
The list of the respective computational problems
includes the solution of a polynomial equation and 
linear and polynomial systems of equations, univariate and multivariate
 polynomial evaluation, interpolation, factorization and decompositions,
rational interpolation, computing matrix factorization and decompositions
(which in turn include various triangular and orthogonal factorizations such as LU, PLU,
QR, QRP, QLP, CS, LR, Cholesky factorizations and eigenvalue and singular value 
decompositions), computation of the matrix
 characteristic and minimal polynomials, determinants, Smith and 
Frobenius normal forms, ranks, and (generalized) inverses, 
univariate and multivariate polynomial resultants, Newton's polytopes, 
greatest common divisors, and least common multiples as well as
manipulation with truncated series and algebraic sets.

Such problems can be solved
by using the error-free symbolic 
computations with infinite precision. 
Computer algebra systems such as Maple and Mathematica compute the solutions 
based on various nontrivial computational techniques such as modular computations,
the Euclidean algorithm and continuous fraction 
approximation, Hensel's and Newton's lifting, Chinese Remainder algorithm,
elimination and resultant methods, and Gr\"obner 
bases computation. The price to achieve perfect accuracy is the substantial  
memory space and computer time required to support the computations.  

The alternative numerical methods rely on operations with binary or decimal numbers 
truncated or rounded to a fixed precision. Operating with the IEEE standard 
floating point numbers represented with single or double precision enables much 
faster computations using much smaller memory but requires
 theoretical and/or experimental study of the impact of rounding errors
on the output. The study involves forward and backward error analysis, linear
and nonlinear operators, and
advanced techniques from approximation and perturbation theories.
Solution of some problems involves more costly computations with 
extended precision.
The resulting algorithms 
support high performance libraries and packages of subroutines such as
those in  
Matlab, NAG SMP, LAPACK, ScaLAPACK, ARPACK, PARPACK, MPSolve, and EigenSolve.


In this chapter we cover both approaches, whose combination 
frequently increases their power
 and enables more effective computations. We focus on the
algebraic algorithms in the large, popular and highly important 
fields of matrix computations and root-finding for univariate polynomials 
and systems of multivariate polynomials. We cover part of these 
huge subjects and
include basic 
bibliography for further study.  
To meet space limitation we
cite books, surveys, and comprehensive articles with pointers to
further references, rather than including all the original technical
papers. Our expositions in Sections~2 and~3 
follow the line of
the first surveys in this area in \cite{16.Pa84:1984speed, 16.Pa92,
  16.Pa97:1997survey, 16.Pa98, 16.Pan98}.

We state the complexity bounds under the random access machine (RAM)
model of computation~\cite{16.AHU74:1974,  16.GaGe03}. 
In most cases we assume the
{\it arithmetic} model, that is we assign a unit cost to addition, 
subtraction, multiplication, and division of real numbers, as 
well as to reading or writing them into a memory location. This model is 
realistic for computations with a fixed (e.g., the IEEE standard 
single or double) precision,
which fits the size of a computer word,  and then the arithmetic model 
turns into the {\em word} model \cite{16.GaGe03}. 
In other cases we
allow working with 
extended precision and assume the {\it Boolean} or {\it bit} model, 
assigning a unit cost to every Boolean or bitwise operation. This
accounts for both arithmetic operations and the length (precision) of the 
operands. We denote the bounds for this complexity by $\OB(\cdot)$.
We explicitly specify whether we use the arithmetic, word, or Boolean model 
unless this is clear from the context. 

We write {\em ops} for ``arithmetic operations", 
 ``$\log$" for ``$\log_2$"
unless specified otherwise, and $\tilde \OB(\cdot)$ 
to show that we are ignoring logarithmic factors.


\section[Matrix Computations 
\protect\chaptocbreak]{Matrix Computations 
}\label{16.sec2}


Matrix computations is the most popular and a highly important area of scientific and
engineering computing. Most frequently they are performed numerically,
with
values represented using  
the IEEE standard single or double precision.  
%
In the chapter of this size we must omit or just barely touch on many important subjects
of this field. 
The reader can find further material and bibliography 
in the surveys~\cite{16.Pa84:1984speed,
               16.Pa92}
and the
books~\cite{16.BDDR00,
	    16.BBC93,
	    16.BiPa94:1994algorithms,
	    16.Dem97,
	    16.DDSV98,
	    16.DER86,
            16.GoLo96,
            16.High02,
	    16.Pan01,
	    16.Ste98,	
            16.TB97,
	    16.Wi65:1965eigenvalue}
and for more specific subject areas in
\cite{16.BDDR00,
            16.GoLo96,
            16.Ste98,
            16.TB97,
            16.Wat07,
	    16.Wi65:1965eigenvalue}
on eigendecomposition and SVD, 
\cite{16.BBC93,
	    16.Dem97,
	    16.DDSV98,
	    16.GoLo96,
            16.High02,
	    16.Ste98,	
	    16.TB97}
on other numerical matrix factorizations, \cite{16.Bjo96, 16.LaHa74}  
on the over- and
under-determined linear systems, their  least-squares
solution, and various other numerical computations with singular matrices, 
\cite{16.HaMaTr11} on randomized matrix computations,
\cite{16.KS99, 16.Pan01} on structured matrix computations,
\cite{16.BiPa94:1994algorithms, 16.GoLo96,  16.Pan96corr, 16.Qu94} 
on parallel matrix algorithms, and \cite{16.ChSt05, 16.Co94, 16.DGP02, 16.DGP04, 
16.GaGe03, 16.GeCzLa92:1992Algebra, 16.KaPa91:1991spaa, 16.KaPa92:1992focs,
16.Pa93, 16.Pan96corr, 16.PaRe93, 
16.Pan11, 16.Tar81, 16.Wie86} 
on ``Error-free Rational Matrix Computations", including computations
over
finite fields, rings, and semirings that produce solutions to linear 
systems of equations, matrix inverses, ranks, determinants, characteristic 
and minimal polynomials, and Smith and Frobenius normal forms. 


\subsection{Dense, Sparse and Structured Matrices. \\
Their Storage and Multiplication by Vectors}


An $m\times n$ matrix $A=[\ a_{i,j}\ ,\ i=1,\ldots,m;\ j=1,\ldots,n\ ]$ is
also denoted $[a_{i,j}]_{i,j=1}^{m,n}$ and $[{\bf A}_1~|~\dots~|~{\bf A}_{m}]$;
it is
a 2-dimensional array with the $(i,j)$th entry $[A]_{i,j}=a_{i,j}$ and 
the $j$th column ${\bf A}_j$. 
$A^T$ is the transpose of $A$. 
Matrix $A$ is a column vector   
if $n=1$ and
 a row vector if $m=1$.
Vector ${\bf v}=[v_i]_{i=1}^{n}$ is an $n$th dimensional column vector. 
The straightforward algorithm 
computes 
the product $A{\bf v}$ by performing  $(2n-1)m$ ops; this is  
optimal for general 
(dense unstructured) $m\times n$ matrices, represented with 
their entries,
but numerous applications  involve structured   matrices
represented with much fewer than $mn$ 
scalar values. 
A matrix is singular if its product by some vectors vanish;
they form its {\em null space}.

An  $m\times n$ matrix is {\em sparse} if it is filled mostly with zeros, 
having only $\phi=o(mn)$ nonzero entries. An important class is
the matrices associated with graphs that 
have {\em families of small separators} \cite{16.GiTa87:1987nested, 16.LiRoTa79}.
This includes {\em banded} matrices $[b_{i,j}]_{i,j}$ with 
small {\em bandwidth} $2w+1$ such that $b_{i,j}=0$ unless $|i-j|\le w$.
A sparse matrix can be stored economically by using appropriate data structures
and can be multiplied by a vector fast, in $2\phi-m$ ops.
Sparse matrices arise in many important applications,
e.g., to solving ordinary and partial
differential equations (ODEs and PDEs) and graph computations. 

Dense structured $n\times n$ 
matrices are usually defined by $O(n)$ parameters,
and one can apply FFT to multiply such matrix by a
vector by using $O(n\log n)$ or $O(n\log^2 n)$ ops 
\cite{16.Pan01}.
Such matrices are omnipresent in applications in signal
and image processing, coding, ODEs, PDEs,
 particle simulation, and Markov chains. 
Most popular among them are the {\it Toeplitz matrices} $T=[t_{i,j}]_{i,j=1}^{m,n}$
and the
{\it Hankel matrices} $H=[h_{i,j}]_{i,j=1}^{m,n}$ where
$t_{i,j}=t_{i+1,j+1}$ and $h_{i,j}=h_{i+1,j-1}$ for all $i$ and $j$
in the range of their definition. Each such matrix is
defined by $m+n-1$ entries of
its first row and first or last column. Products
 $T{\bf v}$ and $H{\bf v}$  can be equivalently written as 
polynomial products or vector convolutions;
their FFT-based computation takes
 $O((m+n)\log (m+n))$ ops per product
\cite{16.AHU74:1974, 16.BiPa94:1994algorithms,
16.Pan01}.
Many other fundamental computations with Toeplitz and other structured matrices 
can be linked to polynomial computations 
enabling acceleration in both areas of computing \cite{16.BGP02/04, 16.BGP03/05, 
16.BGP04, 
16.BiPa86, 
16.BiPa94:1994algorithms, 
16.BoMoPa98/00, 16.EmiMou99jsc, 
16.EmPa02, 16.MoPa98, 16.MoPa00, 16.MoPaRu02/03, 16.Pa92, 16.Pan96corr, 16.Pan01, 16.PIMRTW,
16.PZ10/11, 16.PZ11}.
Similar properties hold for Van\-der\-monde matrices $V=[v_{i}^j]_{i,j=0}^{m-1,n-1}$
and Cauchy matrices $C=[\frac{1}{s_i-t_j}_{i,j}]_{i,j=1}^{m,n}$
where $s_i$ and $t_j$ denote $m+n$ distinct scalars.

One can extend the structures of {\it Hankel}, 
{\it B\'{e}zout,} {\it Sylvester,} {\em Frobenius (companion)}, {\it Vandermonde,} 
and {\it Cauchy} matrices to more general classes of matrices
by associating linear displacement operators. 
(See \cite{16.BiPa94:1994algorithms, 16.Pan01}
 for the details and the bibliography.)
The important classes of {\em semiseparable},
{\em quasiseparable} and other 
{\em rank structured} $m\times n$ matrices generalize 
banded matrices and their inverses; they are expressed by $O(m+n)$ parameters and can be
multiplied by vectors by
performing $O(m+n)$
ops \cite{16.EG99, 16.VVM07}.


\subsection{Matrix Multiplication, Factorization, Randomization}


The straightforward algorithm computes the $m\times p$ product $ AB $
of $m\times n$ by $n\times p$ matrices by using $2mnp-mp$ ops, which is $2n^3-n^2$
if $m=n=p$. 
This upper bound is not sharp. 
Strassen 
decreased it to $O(n^{2.81})$ ops in 1969.
His result was first improved in \cite{16.Pa78} and 10 times afterward,
most recently
by
Coppersmith and Winograd in \cite{16.CoWi90:1990},
Stothers in \cite{16.Stot10}, and Vasilevska Williams in \cite{16.VW}, 
who use $Cn^{\omega}$ ops for $\omega<2.376$, $\omega<2.374$ and
$\omega<2.3727$, respectively. Due to the huge overhead 
constants $C$, however, we have that $Cn^{\omega}<2n^3$
only for enormous values $n$. 
The well recognized group-theoretic techniques \cite{16.CKSU}
enable a distinct description of the known matrix multiplication algorithms, 
but so far have only
supported the same upper bounds on the complexity as the preceding works.
References \cite{16.Stot10} and \cite{16.VW}
extend the algorithms given in Reference
\cite{16.CoWi90:1990}, which in turn combines  
 arithmetic progression technique with the previous advanced  techniques.
Each technique, however, contributes to a dramatic increase of
the overhead constant that makes the resulting algorithms 
practically noncompetitive. 

The only exception is the {\em trilinear aggregating} 
 technique of \cite{16.Pa72}
(cf. \cite{16.Pa84:1984speed}),  
which alone supports the exponent 2.7753 \cite{16.LPS92:1992}
and together with the Any Precision Approximation (APA)
techniques of 
\cite{16.Pa84:1984speed} was an indispensable ingredient
of all algorithms that have beaten Strassen's exponent~2.81 of~1969.
The triple product property (TPP), which is the basis 
of \cite{16.CKSU}, may very well have a natural link to 
trilinear aggregating, although the descriptions available
for the two approaches are distinct.
For matrices of  realistic sizes the numerical algorithms 
in  \cite{16.Kap04:2004}, relying  on 
trilinear aggregating,
use about as many ops as the 
 algorithms of Strassen of 1969 and Winograd of 1971 but  
need substantially less memory space and are more stable numerically.

The
exponent $\omega$ of matrix multiplication
is fundamental for the theory of computing
because $O(n^{\omega})$ or $O(n^{\omega}\log n)$ bounds the complexity of
many important matrix computations such as the computation of 
 $\det\ A$, the {\em determinant}
of an $n\times n$ matrix $A$; its {\it inverse} $A^{-1}$ (where
$\det\ A\ne 0$); 
its {\em characteristic polynomial} $c_A(x)=\det(\ xI-A\ )$
and {\em minimal polynomial} $m_A(x)$, for a scalar variable $x$; 
the Smith and Frobenius normal forms;
the {\it rank,} rank $A$; a submatrix of $A$ having the maximal rank,
the solution vector ${\bf x}=A^{-1}\ {\bf v}$ to a nonsingular
{\it linear system of equations} $A\ {\bf x}={\bf v}$, and
various {\it orthogonal} and {\it triangular factorizations}
of the matrix $A$, as well as various
{\it computations with singular matrices} and seemingly
unrelated combinatorial and graph computations, e.g.,
pattern recognition or
 computing 
all pair shortest distances in a graph~\cite[p. 222]{16.BiPa94:1994algorithms} or 
its transitive closure ~\cite{16.AHU74:1974}. Consequently, all these operations use 
$O(n^{\omega})$ ops where theoretically $\omega<2.3727$
\cite[chap.6]{16.AHU74:1974},~\cite[chap. 2]{16.BiPa94:1994algorithms}.
In practice, however, the solution of all these problems
takes order of $n^3$ ops,
because of the huge overhead constant $C$  of
 all known algorithms that multiply $n\times n$ matrices in $Cn^{\omega}$
ops for $\omega<2.775$, the overhead of the reduction
to a matrix multiplication problem,
 the memory space requirements, and numerical stability problems  \cite{16.GoLo96}.

Moreover, 
the straightforward algorithm 
for matrix multiplication 
remains the users' 
choice  because it is highly effective 
on parallel and pipeline
architectures
 \cite{16.GoLo96, 16.Qu94}; on many computers it supersedes even
the so called ``superfast" algorithms, which  multiply a pair of $n\times n$ 
structured matrices in nearly 
linear arithmetic time,
namely, by using $O(n \log n)$ or $O(n\log^2n)$ ops, where both input and output
matrices are represented with their short generator matrices having
$O(n)$ entries \cite{16.Pan01}. 

Numerous important practical
problems have been reduced to matrix multiplication
because it is so effective. This has also motivated 
the development of block matrix algorithms (called {\em level-three BLAS},
which is the acronym for Basic Linear Algebra Subprograms).

Devising asymptotically fast matrix multipliers,
however, had independent technical interest.
E.g., trilinear aggregating was a nontrivial decomposition of 
the 3-di\-men\-sional tensor associated with matrix multiplication,
and \cite{16.Pa72} was the first of now numerous examples where 
nontrivial tensor decompositions enable dramatic acceleration of important
matrix computations \cite{16.KoBa09, 16.MMD08, 16.OT09}.





The two basic techniques below 
extend matrix multiplication. Hereafter $O$ denotes  matrices 
filled with zeros; $I$ is the square
identity matrices, with ones on the diagonal and zeros elsewhere.

Suppose we seek the  {\it Krylov sequence}
or {\it Krylov matrix} $[B^i {\bf v}]_{i=0}^{k-1}$
for an $n\times n$ matrix $B$ and an $n$-dimensional
vector ${\bf v}$ \cite{16.GoLo96, 16.Green97, 
16.Wie86};
in block Krylov computations 
the vector ${\bf v}$ is replaced by a matrix.
The straightforward algorithm uses
$(2n-1)n(k-1)$ ops, that is about $2n^3$ for $k=n$. An alternative algorithm 
first computes the matrix powers

$$    B^2, B^4, B^8, \ldots, B^{2^s}\ ,\qquad s =
      \lceil\ \log k \ \rceil-1\ ,
$$
and then the products of
$n\times n$ matrices $B^{2^i}$ by $n\times 2^i$ matrices,
for $i=0,1,\ldots,s$:
\begin{eqnarray*}
  & B & {\bf v}\ ,                                                    \\
  & B^2 & [\ {\bf v},\ B{\bf v}\ ] = \left[\ B^2{\bf v},\ B^3{\bf v}\ \right]\ ,        \\
  & B^4 & \left[\ {\bf v},\ B{\bf v},\ B^2{\bf v},\ B^3{\bf v}\ \right] =
     \left[ \ B^4{\bf v},\ B^5{\bf v},\ B^6{\bf v},\ B^7{\bf v}\ \right]\ ,             \\
  & \vdots &
\end{eqnarray*}
The last step completes the evaluation of the Krylov sequence in
$2s + 1$ matrix multiplications, by using $O(n^{\omega} \log k)$ ops overall.

Special techniques for parallel computation
of Krylov sequences for sparse and/or structured matrices
$A$ can be found in~\cite{16.Pa95:1995krylov}. According to these
techniques, Krylov sequence is recovered from the solution to the
associated linear system $(I-A)\ {\bf x}={\bf v} $, which is
solved fast in the case of a special matrix $A$.

Another basic idea of matrix algorithms is to represent the input matrix $A$ as a block
matrix and to operate with its blocks rather than with its
entries. E.g., one can compute $\det A$ and $A^{-1}$ by first
factorizing $A$ as a $2\times 2$ block matrix,
\begin{equation}\label{eqbl}
  A = \left[ \begin{array}{cc}
              I                    & O \cr
              A_{1,0} A_{0,0}^{-1} & I \end{array} \right]
      \left[ \begin{array}{cc}
              A_{0,0} & O \cr
              O       & S \end{array} \right]
      \left[ \begin{array}{cc}
              I & A_{0,0}^{-1} A_{0,1} \cr
              O & I \end{array} \right]~
\end{equation}
where $S=A_{1,1} - A_{1,0} A_{0,0}^{-1} A_{0,1}$. 
The $2\times 2$ block triangular factors are readily invertible,
$\det A=(\det A_{0,0})\det S$ and $(BCD)^{-1}=D^{-1}C^{-1}B^{-1}$, 
and so the cited tasks for the input $A$ are reduced to the same tasks 
for the half-size matrices $A_{0,0}$
and $S$. It remains to factorize them recursively. 
The northwestern blocks (such as $A_{0,0}$),
called leading principal submatrices, must be nonsingular throughout
the recursive process, but this property holds for the 
highly important class of {\em symmetric positive definite} matrices $A=C^TC$, 
$\det C\neq 0$, and
can be also achieved by means of symmetrization,
pivoting, or randomization ~\cite[chap. 6]{16.AHU74:1974}, 
\cite[chap. 2]{16.BiPa94:1994algorithms},
~\cite[sects. 5.5 and 5.6]{16.Pan01}).
Recursive application of (\ref{eqbl}) should produce the LDU factorization 
$A=LDU$ where the matrices $L$ and $U^T$ are lower triangular and $D$ diagonal.
Having this factorization computed , we can readily solve linear systems
$A{\bf x}_i={\bf b}_i$ for various vectors ${\bf b}_i$, by using about
$2n^2$ ops for each $i$, rather than $\frac{2}{3}n^3+O(n^2)$ in Gaussian 
elimination. 

{\em Factorizations} (including PLU,
QR, QRP, QLP, CS, LR, Cholesky factorizations and eigenvalue and singular value 
decompositions)
are the most basic tool of matrix computations (see, e.g., \cite{16.Ste98}),
recently made even more powerful with {\em randomization} (see \cite{16.HaMaTr11,
16.PGMQ, 16.PIMRTY, 16.PaQi, 16.PaQia, 16.PQZ11, 16.PQZa,
 16.PQZc},
and the bibliography therein).
It is well known that random matrices tend to be nonsingular and
well conditioned  (see, e.g., \cite{16.SST06}),
that is they lie far from singular matrices 
and therefore \cite{16.GoLo96, 16.High02, 16.Ste98}
are not sensitive to rounding errors and are 
suitable for numerical computations.
The solution ${\bf x}=A^{-1}{\bf b}$ of
a nonsingular linear system  $A{\bf x}={\bf b}$
of $n$ equations can be obtained with a precision $p_{\rm out}$
in $O_{\tilde B}(n^3p+n^2p_{\rm out})$ Boolean time 
for a fixed low precision $p$ 
provided the matrix $A$ is well conditioned;
that accelerates Gaussian elimination by an order of magnitude for large $n+p_{\rm out}$.
Recent randomization techniques in \cite{16.HaMaTr11,
16.PGMQ, 16.PIMRTY, 16.PaQi, 16.PaQia, 16.PQZ11, 16.PQZa,
 16.PQZc} extend this property to much larger
class of linear systems and
enhance the power 
of various other matrix computations with singular or ill conditioned matrices,
e.g., their approximation by low-rank matrices,
computing a basis for the null space of a singular matrix,
and approximating such bases for nearby singular matrices. 
Similar results have been proved for rectangular and Toeplitz matrices. 

We refer the reader to  \cite{16.HaMaTr11, 16.SST06, 16.GI10}
on impressive progress achieved in many other areas of matrix computations
by means of randomization techniques.

\subsection{Solution of linear systems of equations}


The solution of a linear system of $n$ equations,
$A{\mathbf{ x}}={\mathbf{ b}}$ is the most frequent operation in
scientific and engineering
computations and is highly important theoretically.
Gaussian elimination solves such a 
system by applying 
 $(2/3)n^3+O(n^2)$ ops.

Both Gaussian elimination and {\em (Block)
Cyclic Reduction} use $O(nw^2)$ ops 
for banded linear systems with bandwidth $O(w)$. 
One can solve rank structured linear systems in $O(n)$ ops 
\cite{16.EG99, 16.VVM07}; generalized nested dissection\
uses $O(n^{1.5})$ flops for the inputs associated with small
separator families \cite{16.LiRoTa79, 16.Pa93, 
16.PaRe93}.

Likewise, we can dramatically accelerate Gaussian elimination for
 dense structured input matrices represented with their short generators,
defined by the associated {\em displacement operators}. This includes Toeplitz, Hankel,
Vandermonde, and Cauchy matrices as well as matrices with similar structures. 
The MBA divide-and-conquer ``superfast" algorithm (due to papers by Morf of 1974/1980
and Bitmead and Anderson of 1980) solves nonsingular structured linear systems 
of $n$ equations in 
$O(n \log ^2 n)$ ops by applying the
recursive $2\times 2$ block factorization (\ref{eqbl})
and preserving matrix structure
\cite{16.BiPa94:1994algorithms, 16.Pan01, 16.PMRa, 16.PaWa08}.
In the presence of rounding errors, however,
Gaussian elimination,
the MBA  and Cyclic Reduction
algorithms easily fail unless one applies 
pivoting, that is interchanges the equations (and sometimes 
unknowns) to avoid divisions by absolutely small numbers. A by-product 
is the factorization $A=PLU$ or $A=PLUP'$, for lower triangular matrices
$L$ and $U^T$ and permutation matrices $P$ and $P'$.

Pivoting, however, takes its toll.
It 
``usually degrades the performance" \cite[page 119]{16.GoLo96} 
by interrupting the string of arithmetic computations with the foreign operations 
of comparisons, is not friendly to 
block matrix algorithms and updating input matrices,
hinders parallel processing and pipelining, and 
tends to destroy structure and sparseness,
except for the inputs that have
Cauchy-like and Van\-der\-monde-like
structure. 
The latter exceptional classes have been extended
to the inputs with 
structures of Toep\-litz/Han\-kel type
by means of {\em displacement transformation}  \cite{16.Pa90, 16.Pan01}. 
The users welcome this numerical stabilization, even though
it slows down the MBA algorithm by a factor of $n/\log^2 n$,
that is from ``superfast" to ``fast", which is still
by a factor of $n$ faster than the solution for general unstructured inputs,
which takes order $n^3$ ops.

Can we avoid pivoting in numerical
algorithms with rounding for general, sparse and structured linear systems
to achieve both numerical stability and superfast performance?
Yes, for the important classes  
where the  input matrices $A=(a_{ij})_{i,j}$
are
diagonally dominant, that is $|a_{ii}|>\sum_{i\neq j}|a_{ij}|$
or $|a_{ii}|>\sum_{j\neq i}|a_{ij}|$ for all $i$,
or
symmetric positive definite,
that is $A=C^TC$ for a nonsingular matrix $C$.
To these input classes
 Gaussian elimination,
Cyclic Reduction, and the
 MBA algorithm can be safely applied 
with rounding and with no pivoting. 
For some other classes of sparse and positive definite linear systems, 
pivoting has been modified 
into nested dissection, Markowitz heuristic rule, and other techniques that
preserve sparseness during the elimination yielding faster solution without
causing numerical problems   
\cite{16.DDSV98,
16.GiSc92, 
16.LiRoTa79, 16.Pa93, 
16.PaRe93}.  
Can we extend these benefits to other input matrix classes?

Every nonsingular linear system $A\ {\bf x}={\bf b}$
is equivalent to the symmetric positive definite ones
$A^TA\ {\bf x}=A^T{\bf b}$
and $A\ A^T{\bf y}={\bf b}$ where ${\bf x}=A{\bf y}$,
but great caution is recommended in such symmetrizations because 
the condition number $\kappa (A)=||A||_2||A^{-1}||_2\ge 1$
is squared in the transition to the matrices $A^TA$ and $AA^T$,
which means growing propagation and magnification of rounding
errors. 

There are two superior directions.
The algorithms of \cite{16.PQZ11, 16.PQZa, 16.PQZC} avoid pivoting
for general and structured linear systems 
by applying randomization. These techniques are recent,
proposed in \cite[Sect.~12.2]{16.PGMQ},
but their effectiveness has formal and 
experimental support.  

A popular classical alternative to Gaussian elimination is the iterative solution, 
e.g., by means of the Conjugate 
Gradient and GMRES algorithms \cite{16.Ben02, 16.GoLo96, 16.Green97, 
16.vdV03}. They compute sufficiently long Krylov sequences 
(defined in the previous section) and then approximate the solution with linear
combinations $\sum_i c_iA^i{\bf b}$ or $\sum_i c_i(A^TA)^iA^T{\bf b}$ 
for 
proper coefficients $c_i$.
The cost  of computing the product of the matrix $A$ or $A^TA$ by a 
vector is  dominant, but it is small for structured and sparse matrices $A$. 
One can even
call a matrix sparse or structured if and only if it can be multiplied by a 
vector fast. 


Fast convergence to the solution is critical.
It is not generally guaranteed but proved for 
some important classes of input matrices. 
The major challenge
are the extension of these classes and the design of powerful
methods for special input classes, notably 
{\em multilevel methods} (based on the 
{\em algebraic multigrid})
~\cite{16.MiPa80, 16.Mc87:1987multigrid, 16.PaRe92:1992multigrid}
and tensor decompositions \cite{16.OT09,16.KoBa09},
 highly effective for many linear systems arising in
discretization of ODEs, PDEs,  and integral equations. 

{\em Preconditioning} of the input matrices 
at a low computational cost accelerates convergence of iterations 
 for many important classes of sparse and structured linear systems 
\cite{16.Ben02, 
16.Green97}, and more recently, 
based on randomized preconditioning, for quite
 general as well as structured linear systems 
 \cite{16.PGMQ, 16.PIMRTY, 16.PaQi,  16.PaQia, 16.PQZ11, 16.PQZa, 
 16.PQZc}. 

One can iteratively 
approximate the inverse or pseudo-inverse of a matrix
\cite[Section 5.5.4]{16.GoLo96} by means of
Newton's iteration
$X_{i+1}=2X_i-X_iMX_i$, $i=0,1,\dots$.
We have $I-MX_{i+1}=(I-MX_i)^2=(I-MX_0)^{2^{i+1}}$;  
therefore, the residual norm $||I-MX_{i}||$
is  squared 
in every iteration step,
$||I-MX_{i}||\le ||I-MX_0||^{2^{i}}$ for $i=1,2,\dots$,
and so convergence is very fast
unless $||I-MX_0||\ge 1$ or is near 1.
The cost of two matrix multiplications is dominant per an iteration step;  
this makes the computation fast on multiprocessors
as well as in the case of structured matrices $M$ and $X_i$.
See more on
 Newton's iteration, including the study of its initialization, convergence, 
and preserving displacement matrix structure, in  \cite[chapters 4 and 6]{16.Pan01}, 
\cite{16.PS91, 16.PBRZ99, 16.PRW02, 
16.PaWa03, 
16.PKRK,
16.Pan10}.


\subsection{Symbolic Matrix Computations}


Rational matrix computations for a rational or integer input (such as the solution 
of a linear system and computing the determinant of a matrix) can be performed with no errors. 
To decrease the computational cost, one should control the growth of the precision 
of computing. 
Some special techniques 
achieve this in rational Gaussian elimination
~\cite{16.Bar68:1968exactdivision, 16.GeCzLa92:1992Algebra}. As a more fundamental
tool one can reduce the computations modulo 
a sufficiently large integer $m$ to obtain
 the rational or 
integer output values $z=p/q$ (e.g., the solution vector for a linear system) 
 modulo $m$. Then we can
 recover  $z$ from two integers $m$ and $z$ mod $m$ 
by applying
the continued fraction approximation
algorithm,  in other contexts  called Euclidean algorithm
\cite{16.GaGe03, 16.WaPa03}.
Instead we can readily obtain
$z=z\ {\rm mod}\ m$ if $ z~{\rm mod}~m <r$
or $z=-m+z\ {\rm mod}\ m$ if $ z~{\rm mod}~m <r$ otherwise,
provided we know that the integer $z$ lies in the range
$[-r,r]$ and if $m>2r$.
 
Computing the determinant
of an integer matrix, we can 
choose the modulus $m$ based on Hadamard's bound. 
A nonsingular linear system
$A {\bf x}={\bf v}$ can become singular 
after the reduction modulo a prime $p$ but
only with a low probability for a random choice of a
prime $p$ in a fixed sufficiently large interval as well as 
for a reasonably large power of two and a random integer matrix
\cite{16.PaWa08}.

One can choose $m=m_1m_2 \cdots m_k$ 
for pairwise relatively prime integers 
$m_1,m_2,\ldots, m_k$ (we call them {\em coprimes}),
then compute $z$ modulo all these coprimes,
 and finally recover $z$ by applying the Chinese Remainder algorithm 
\cite{16.AHU74:1974, 16.GaGe03}.  
The error-free computations modulo
$m_i$ require the precision of $\log m_i$ bits; 
the cost of computing the values $z$ mod $m_i$
for $i=1,\dots,k$ dominates
the cost of the subsequent recovery of the value 
$z$ mod $m$.

Alternatively one can apply 
$p$-$adic$ ({\it Newton--Hensel}\/) {\it lifting} \cite{16.GaGe03}.
For solving linear systems of equations and matrix inversion
they can be viewed as the
symbolic counterparts to    
iterative refinement
and Newton's iteration of the previous section, 
both well known in numerical linear algebra
\cite{16.Pan11}.

Newton's lifting begins with a prime $p$, a larger integer $k$,
 an integer matrix $M$, and its
inverse $Q=M^{-1} \bmod p$, such that $I-QM \bmod p=0$. Then one writes $X_0=Q$,
recursively computes the matrices $X_{j}=2X_{j-1}-X_{j-1}MX_{j-1} \bmod (p^{2^j})$,
notes that $I-X_jM=0 \bmod (p^{2^j})$ for $j=1,2,\dots,k$,
and finally recovers the inverse matrix $M^{-1}$ from $X_k=M^{-1}$ mod $p^{2^k}$.

Hensel's lifting begins with the same input complemented with an integer vector $\bf b$.
Then one writes 
${\bf r}^{(0)}={\bf b}$, recursively computes the vectors
         $${\bf u}^{(i)} = Q {\bf r}^{(i)}\bmod p, ~~
         {\bf r}^{(i+1)}=({\bf r}^{(i)}-M{\bf u}^{(i)})/p,~~i= 0,1,\ldots,k-1,$$
         and ${\bf x}^{(k)}=\sum^{k-1}_{i=0}{\bf u}^{(i)}p^i$
such that  $M {\bf x}^{(k)} = {\bf b}\bmod (p^k)$,
and finally recovers the solution ${\bf x}$ to the linear system $M{\bf x}={\bf b}$
from the vector  ${\bf x}^{(k)}={\bf x} \bmod (p^k)$.

Newton's and Hensel's lifting are  
particularly powerful where the input matrices $M$ and $M^{-1}$ 
are sparse and/or structured, e.g.,
Toeplitz, Hankel, Vandermonde, Cauchy. 
Hensel's 
lifting 
enables the solution in nearly optimal time under both Boolean and word models
\cite{16.Pan11}. 
We can choose $p$ being a
power of two and use  computations in the  
 binary mode. 
Reference~\cite{16.EGGSV07} discusses lifting for sparse linear systems.


\subsection{Computing the Sign and the Value of a Determinant}


The value or just the sign of $\det A$,
the determinant of a square matrix $A$, are required
in some fundamental geometric and algebraic/geometric computations 
such as the computation of convex hulls, 
Voronoi diagrams, algebraic curves and surfaces, multivariate and univariate
resultants and Newton's polytopes. Faster numerical methods are preferred 
as long as the correctness of the output can be certified.
In the customary {\em arithmetic 
filtering} approach,  one applies fast numerical methods as long as they work
and, in the rare cases when they fail, shifts to the slower symbolic methods. 
For fast numerical computation of $\det A$ one can employ factorizations $A=PLUP'$
(see Section 2.2) or $A=QR$ \cite{16.Cla92, 16.GoLo96}, 
precondition 
the matrix $A$  \cite{16.PGMQ}, and then
certify the output 
sign  \cite{16.PaYu99/01}. 

If $A$ is a rational or integer matrix, then 
 the Chinese Remainder algorithm
of the previous subsection is highly effective, 
particularly using heuristics for working modulo $m$ for $m$ much smaller than 
Hadamard's  
bound on $|\det A|$  \cite{16.BEPP97/99}. 

Alternatively
\cite{16.Pa87, 16.Pa88, 
16.EGV00},
one can solve linear systems $A{\bf y}(i)={\bf b}(i)$ for random vectors ${\bf b}(i)$
and then apply Hensel's lifting to recover 
$\det A$ as a least common denominator of the rational components
of all ${\bf y}(i)$.

Storjohann in \cite{
16.Sto05} advanced randomized Newton's lifting to 
yield $\det A$ more directly in the optimal asymptotic Boolean time $\OB(n^{\omega+1})$ for
$\omega <2.3727$.
Wiedemann in 1986, Coppersmith in 1994, and a number of their successors
compute $\det A$ by 
extending the Lanczos and block Lanczos classical algorithms.
This is particularly effective for sparse or structured matrices $A$
and in further extension to multivariate determinants and resultants
(cf.\ \cite{16.KaVi04, 16.EmPa02, 16.EmPa03/05, 16.Pan04}). 



\section{Polynomial Root-Finding and
Factorization}


\subsection{Computational Complexity Issues}


Approximate solution of an $n$th degree polynomial equation,
\begin{equation}\label{eqpol}
  \displaystyle
    p(x) = \sum_{i=0}^n\ p_i\ x^i=p_n \prod_{j=1}^n(x-z_j)=0\ ,\qquad p_n\ne 0,
\end{equation}
that is the approximation of the roots $z_1,\dots,z_n$ for given coefficients
$p_0,\dots,p_n$, is a classical problem that has greatly influenced the
development of mathematics and computational mathematics 
throughout four millennia, since the Sumerian times
\cite{16.Pa97:1997survey, 16.Pa98}.
The problem remains highly important for the
theory and practice of the present day algebraic and algebraic/geometric computation, 
and 
new root-finding algorithms appear every year \cite{16.McN02, 16.McN07, 
16.McNP12, 16.McNPa}.
 
To approximate even a single root 
of a monic polynomial $p(x)$ within error 
bound $2^{-b}$
 we must process at least $(n+1)nb/2$
bits
 of the input coefficients $p_0,\dots,p_{n-1}$.
Indeed perturb the $x$-free
coefficient of the polynomial $(x-{6/7})^n$  
 by $2^{-bn}$. Then the root
$x=6/7$ jumps by $2^{-b}$, and 
similarly if we perturb the coefficients
$p_i$ by $2^{(i-n)b}$ for $i=1,\ldots,n-1 $.
Thus to ensure the output
precision of $b$ bits, we need an input precision of at least
$(n-i)b$ bits for each coefficient $p_i$, $i=0,1,\ldots,n-1$.
We need at least $\lceil (n+1)nb/4\rceil$ bitwise operations
to process these bits, each operation having at most two input bits. 

It can be surprising, but we can
approximate all $n$ roots within $2^{-b}$ 
by using $bn^2$ Boolean (bit) operations up to a polylogarithmic 
factor for $b$ of order $n\log n$ or higher, that is 
we can approximate all roots about as fast as we write down the input.
We achieve this by applying the {\it divide-and-conquer algorithms} in 
\cite{16.Pa:toappearapprox, 16.Pa97:1997survey, 16.Pan01/02} (see \cite{16.Kir98,
16.NeRe94, 16.Sch82} on the related works). The algorithms first
compute a sufficiently wide root-free annulus $A$ on the complex plane, 
whose exterior and interior contain comparable numbers of the roots, 
that is the same numbers up to a fixed constant factor. 
Then the two factors of $p(x)$
are numerically computed, that is $F(x)$, having all its roots in
the interior of the annulus, and $G(x)=p(x)/F(x)$, having no roots there.
Then the polynomials
$F(x)$ and $G(x)$ are recursively factorized until
factorization of $p(x)$ into the product of linear factors is
computed numerically. From this factorization, approximations
to all roots of $p(x)$ are obtained. For approximation of a single root
see the competitive algorithms of \cite{16.Pan00a}.

 It is interesting  that, up to polylog factors, 
both lower and upper bounds on the Boolean time decrease to 
$bn$  \cite{16.Pan01/02}
if we only seek the factorization of $p(x)$, 
 that is, if instead of the roots $z_j $, we compute 
 scalars $a_j$ and $b_j$ such that 
\begin{equation}\label{eqfac}
||p(x)-\prod_{j=1}^n(a_jx-c_j)||<2^{-b}||p(x)|| 
\end{equation}
for the polynomial norm  $||\sum_i q_ix^i||=\sum_i|q_i|$.

 The \textit{isolation of the zeros} 
of  a polynomial 
$p(x)$ of (\ref{eqpol}) having integer
coefficients and simple zeros
is
the computation of $n$ disjoint discs,
each containing exactly
one root of $p(x)$. 
This can be a bottleneck stage of root approximation because 
one can contract such discs by performing a few 
subdivisions and then apply
numerical iterations (such as Newton's) 
that would very rapidly approximate the isolated zeros
within a required tolerance.
Reference \cite{16.Pana} yields even faster refinement by extending
the techniques of
\cite{16.Pa:toappearapprox, 16.Pa97:1997survey, 16.Pan01/02}.
 
Based on the classical ``gap theorem"
(recently advanced in \cite{16.EMT}),
Sch{\"o}nhage in \cite[Sect.~20]{16.Sch82} has reduced  
the isolation problem to computing factorization 
(\ref{eqfac})  for
$b=\lceil(2n+1)(l+1+\log(n+1))\rceil$ where  $l$ is
the maximal coefficient length,
that is the minimum integer
 such that $|\Re(p_j)|<2^l$ and $|\Im(p_j)|<2^l$
for $j=0,1,\dots,n$.
Combining the cited algorithms of \cite{16.Pa:toappearapprox, 16.Pa97:1997survey, 16.Pan01/02}
with this reduction yields

\begin{theorem}\label{thisol}
Let polynomial $p(x)$ of~(\ref{eqpol}) have $n$ distinct simple zeros
and integer coefficients in the range $[-2^{\tau},2^{\tau}]$.
Then one can isolate the $n$ zeros of 
$p(x)$ from each other at the Boolean cost $\tilde \OB(n^2\tau)$.
\end{theorem}

The algorithms of \cite{16.Pa:toappearapprox, 16.Pa97:1997survey, 16.Pan01/02}
incorporate the techniques of \cite{16.NeRe94, 16.Sch82}, but
advance them and support
substantially smaller upper bounds on the computational complexity.
In particular these algorithms decrease by a factor of $n$ 
the estimates 
of \cite[Theorems 2.1, 19.2 and 20.1]{16.Sch82} 
on the  Boolean complexity of polynomial factorization,
root approximation and root isolation.



\subsection{Root-Finding via Functional Iterations}


About the same record  complexity estimates for root-finding
would be also supported by 
some {\em functional iteration} algorithms if one assumes their 
convergence rate defined by ample empirical evidence, 
although never proved formally. 
The users   accept
such an evidence instead of the proof and prefer the latter algorithms because 
they are easy to program and have been carefully implemented;
like the algorithms of \cite{16.Pa:toappearapprox, 16.Pa97:1997survey, 16.Pan00a, 16.Pan01/02}
they allow tuning 
the precision of computing to the precision required for every output root,
which is higher for clustered and multiple roots than for
single isolated roots.

For approximating a single root $z$, the current practical champions 
are modifications of {\it Newton's iteration,}
$z(i+1) = z(i) - a(i) p(z(i))/p^\prime(z(i))$, $a(i)$ being the
step-size parameter~\cite{16.M73:1973}, {\it Laguerre's
method}~\cite{16.F81:1981, 16.HPR77}, and the 
{\it Jenkins--Traub algorithm}~\cite{16.JT70:1970}. One can 
deflate the input polynomial via its
numerical division by $x-z$ to extend these algorithms to
approximating a small number of other roots.
If one deflates many roots, the coefficients of the remaining factor can
grow large as, e.g., in the divisor of the polynomial $p(x)=x^{1000}+1$ that
has degree 498 and shares with $p(x)$ all its roots having positive real parts.

For the approximation of all roots, a good option is
the Weierstrass--Durand--Kerner's (hereafter {\em WDK}) algorithm, 
defined by the recurrence
\begin{equation}
      z_j(l+1) = z_j(l) - {p \left( z_j(l) \right)
                 \over p_n \prod_{i \not= j}
                  \left( z_j(l)-z_i(l) \right)}\ ,
                   \quad j=1,\ldots,n,\quad l=0,1,\ldots~.
\label{16.eq:durker}
\end{equation}
It has excellent empirical global convergence.
Reference ~\cite{16.PZ11} links it to polynomial factorization 
and adjusts it 
to approximating a single root in $O(n)$ ops per step.

 
A customary choice of $n$ initial approximations
$z_j(0)$ to the $n$ roots of the polynomial $p(x)$
(see~\cite{16.BiFi00} for a heuristic alternative)
is given by
$z_j(0)=r\ t\exp(2\pi \sqrt{-1}/n)\ ,\quad j=1,\ldots,n$. 
Here $t>1$ is a fixed scalar and  $r$ is an upper bound on the root radius, 
such that all roots $z_j$ lie in the disc
$\{x:~|x|=r\}$
 on the complex plane.
This holds, e.g., for
\begin{equation}
       r = 2 \max_{i<n} \left| p_i/p_n \right|^{\frac{1}{n-i}}~.
\label{16.eq:radius}
\end{equation}
For a fixed
$l$ and for all $j$ the computation in~(\ref{16.eq:durker}) uses
$O(n^2)$ ops. We can use just
 $O(n \log^2 n)$ ops if we apply fast multipoint polynomial evaluation
algorithms based of fast FFT based polynomial division
~\cite{16.AHU74:1974, 16.BiPa94:1994algorithms, 
16.BoMu75:1975, 16.Pan01, 16.PaLaSa92}, 
but then we would face numerical
stability problems. 

As with Newton's, Laguerre's, Jenkins--Traub's algorithms
and the Inverse Power iteration in \cite{16.BGP02/04, 16.PZ10/11},
one can employ   
this variant of the WDK 
to approximate many or all roots of $p(x)$
without deflation. Toward this goal, one can
concurrently apply
the 
algorithm at sufficiently many distinct initial points 
$z_j(0)=r\ t\exp(2\pi \sqrt{-1}/N)\ ,\quad j=1,\ldots,N\ge n$
(on a large circle for large $t$) 
or according to \cite{16.BiFi00}.
The work can be distributed among processors 
that do not need to interact with each other
until they compute the roots.

See
~\cite{16.McN02, 16.McN07, 16.McNP12, 16.McNPa, 16.Pa97:1997survey} 
and references therein
on this and other effective functional iteration algorithms.
Reference ~\cite{16.BiFi00} covers MPSolve, the most effective current root-finding
subroutines, based on Ehrlich--Aberth's algorithm.


\subsection{Matrix Methods for Polynomial Root-Finding}


By cautiously
avoiding numerical problems \cite[Sec.7.4.6]{16.GoLo96},
one can approximate the roots of $p(x)$
 as the eigenvalues of the associated (generalized)
companion matrices, that is matrices having characteristic polynomial $p(x)$.
Then one can employ numerically stable methods 
and the excellent software available for matrix 
computations, such as the QR celebrated algorithm.  
E.g., Matlab's subroutine {\em roots} applies  
it to the companion matrix of a polynomial. 
Fortune in \cite{16.F01/02} and in his root-finding package 
EigenSolve (citing earlier work of 1995 by Malek and Vaillancourt)
apply it to other generalized companion matrices
and update them when the approximations to the roots are improved.  

The algorithms of \cite{16.BGP02/04, 16.BGP03/05,
16.BGP04, 
16.Pan05,
16.BBEGG, 16.VBV10, 16.PZ10/11, 16.PQZb} exploit the structure of
(generalized) companion matrices, e.g., where they
are diagonal plus rank-one (hereafter {\em DPR1}) matrices,
to accelerate the eigenvalue computations. The papers
\cite{16.BGP02/04, 16.PZ10/11} apply and extend the Inverse Power
method \cite[Section 7.6.1]{16.GoLo96}; they exploit
matrix structure, simplify the customary use of
Rayleigh quotients for updating approximate eigenvalues, and apply special preprocessing 
techniques. For both companion and DPR1 inputs the resulting algorithms 
use linear space and linear arithmetic time per iteration step,
enable dramatic
parallel acceleration, and
deflate the input in
$O(n)$ ops; for  
 DPR1 matrices repeated deflation can 
produce
 all $n$ roots with no numerical problems.

The algorithms of \cite{16.BGP03/05, 
16.BGP04, 
16.BBEGG, 16.VBV10} employ the QR algorithm,
but decrease the arithmetic time per iteration step
from quadratic to linear by exploiting the rank 
matrix structure of companion matrices.
Substantial further refinement of these
techniques is required to make them competitive with
MPSolve. See \cite{16.Za} on recent progress.

The papers~\cite{16.Pan05, 16.PQZb} advance Cardinal's  polynomial
root-finders of 1996, based on repeated squaring.
Each squaring is reduced to    
performing a small number of FFTs and thus uses order $n\log n$ ops. 
One can weigh potential advantage of convergence to nonlinear factors of $p(x)$,
representing multiple roots or root clusters,
at the price of increasing the time per step by a factor of $\log n$
versus  the Inverse Power method, advanced for root-finding 
in \cite{16.BGP02/04, 16.PZ10/11}.

\subsection{Extension to Approximate Polynomial GCDs}


Reference ~\cite{16.Pan98/01} combines polynomial root-finders with
algorithms for bipartite matching to  
compute approximate univariate polynomial greatest common divisor
(GCD) of two polynomials, that is, the GCD of the maximum degree
for two polynomials of the same or smaller degrees
lying in the $\epsilon$-neighborhood of the input polynomials for a fixed 
positive $\epsilon$. 
Approximate GCDs are required
in computer vision, 
algebraic geometry, computer modeling, and control. 
For a single example, GCD defines the intersection of two 
algebraic curves defined by the two input polynomials,
and approximate GCD does this under input perturbations
of small norms.
See \cite{16.BB10} on the bibliography on approximate GCDs,
but see \cite{16.Pa90, 16.Pan01, 16.PQZ11} on the  
 structured matrix algorithms involved.


\subsection{Univariate Real Root Isolation and Approximation}\label{sec:univariate_real_solving}


In some algebraic and geometric computations,
the input polynomial $p(x)$ has real coefficients, and
only its real roots must be approximated. 
One of the fastest real root-finders in the
current practice is still MPSolve, which 
uses almost the same running time
for real roots as for all complex roots.
This can be quite vexing, because very frequently  
the real roots
make up only a small fraction of all roots \cite{16.egt-issac-2010}.
Recently, however, the challenge was taken 
in the papers \cite{16.PZ10/11, 16.PQZb}, whose numerical iterations 
are directed to converge to real and nearly real roots. This 
 promises acceleration by a factor of $d/r$ where the input polynomial
has $d$ roots, of which $r$ roots are real or nearly real.
In the rest of this section we cover an alternative direction, that is
real root-finding by means of
isolation of the real roots of a polynomial.

We write  $p(x) = a_d \, x^d + \cdots + a_1 \, x + a_0$,
assume integral coefficients with
the maximum bit size $\tau= 1 + \max_{i \leq d}\{ \lg{|a_i|} \}$, 
and seek isolation of real
roots, that is seek real line intervals
with rational endpoints, each containing exactly one real root.
We may seek also the root's multiplicity.
We assume rational algorithms, that is,
error-free algorithms that operate with rational numbers.

If all roots of $p(x)$ are simple,  
then the minimal distance between them, the 
{\em separation bound}, is at most $b=d^{-(d+2)/2} (d+1)^{(1-d)/2}2^{\tau(1-d)}$,
or roughly $ 2^{-\sO( d \tau)}$ (e.g., \cite{16.m-mca-92}),
and we isolate real roots as soon as we approximate them 
within less than $b/2$.
Effective solution algorithms rely on Continued Fractions (see below),
having highly competitive implementation in \textsc{synaps} \cite{16.mptt-mega-2005,
16.SNC-exp} and its descendant \textsc{realroot}, a package of \textsc{mathemagix}, 
on the Descartes' rule
of signs, and the Sturm or Sturm--Habicht sequences.

\begin{theorem}\label{realsolve}
The rational algorithms discussed in the sequel
isolate all $r$ real roots of $p(x)$ in $\sOB( d^4 \tau^2)$ bitwise ops.
Under certain probability distributions for the coefficients,
they are expected to use $\sOB( d^3 \tau)$ or $\sOB(r d^2 \tau)$.
 \end{theorem}

The bounds exceed those of Theorem~\ref{thisol}, 
but \cite{16.PaT13} has changed this, by closing the gap.
 Moreover rational solvers are heavily in use,
have long and respected history, and are of independent technical interest.  
Most popular
 are the subdivision algorithms,
such as \func{sturm}, \func{descartes} and \func{bernstein}.
By mimicking binary search, they repeatedly subdivide an initial interval that contains all real roots
until every tested interval contains at most one real root. 
They differ in the way of counting the real roots in an interval. 

The algorithm \func{sturm} 
(due to the work by Sturm of 1835, see  \cite{16.GaGe03})
 is the closest to binary search; it produces isolating intervals and 
root multiplicities at the cost $\sOB(d^4 \tau^2)$
\cite{16.Yap:SturmBound:05,16.emt-lncs-2006};
see \cite{16.egt-issac-2010} on the decrease of
the expected cost to $\sOB(r d^2 \tau)$.

The complexity of both algorithms \func{descartes} and \func{bernstein} is $\sOB(
d^4 \tau^2)$ \cite{16.ESY:descartes,16.emt-lncs-2006}.
Both rely on Descartes' rule of
sign, but
the \func{bernstein} algorithm also employs
the
Bernstein basis polynomial representation. See
\cite{16.Vincent, 16.galuzzi98:vincent}
on the theory and history of \func{descartes},
\cite{16.ColAkr:descartes:76,
16.RouZim:solve:03, 16.EigenEtal:bitstream:05, 16.ms-jsc-2010, 16.s-DN-arxiv-11}
on its modern versions, and 
\cite{16.emt-lncs-2006, 16.MoVrYa02} and the references therein
on the \func{bernstein} algorithm.

The Continued Fraction algorithm, \func{cf}, computes the continued fraction 
expansions of the real roots of the polynomial.
The first formulation of the algorithm is due to Vincent.
By Vincent's theorem
repeated transforms $x \mapsto c + \frac{1}{x}$
eventually yield a polynomial with zero or one sign variation
and thus (by Descartes' rule) with zero or resp. one real root in $(0, \infty)$.
In the latter case the inverse transformation 
computes an isolating interval.
Moreover, the $c$'s in the transform correspond
to the partial quotients of the continued fraction
expansion of the real root.  
Variants differ in the way they
compute the partial quotients.

Recent algorithms control the growth of coefficient bit-size
and decrease the bit-complexity from exponential (of Vincent)
to $\sOB( d^3 \tau)$ expected
and $\sOB( d^4 \tau^2)$ worst-case bit complexity.  
See
\cite{16.mr-jsc-2009, 16.sharma-tcs-2008, 16.t-macis-icf-11, 16.te-tcs-2008}
and the references therein on these results,
history and variants of CP algorithms.




\section{Systems of Nonlinear Equations} \label{16.sec:nonlin}

Given a system
$\{p_1(x_1,\ldots,x_n), \ldots, p_r(x_1,\ldots,x_n)\}$
of nonlinear polynomials with rational coefficients, 
the $n$-tuple of complex numbers $(a_1,\ldots,a_n)$
is a solution of the system if $p_i(a_1,\ldots,a_n) = 0$, 
$1\le i \le r$. 
Each
$p_i(x_1,\ldots,x_n)$ is said to be an element of $\RR{Q}[x_1,\ldots,x_n]$,
the ring of polynomials in $x_1,\ldots,x_n$ over the field of
rational numbers.
In this section, we explore the problem of 
solving a well-constrained system of nonlinear equations, namely when $r=n$,
which is the typical case in applications. 
We also indicate how an initial phase of exact algebraic computation leads
to certain numerical methods that can approximate all solutions;
the interaction of symbolic and numeric computation is currently an
active domain of research, e.g.~ \cite{16.BiPaVe08, 16.EmMoPa04, 16.KMP11}.
We provide an overview and cite references to different symbolic
techniques used for solving systems of algebraic (polynomial) equations.
In particular, we describe methods involving {\it resultant} and
{\it Gr\"{o}bner basis} computations.

{\bf Resultants}, as explained below, formally express the solvability of
algebraic systems with $r=n+1$; solving a well-constrained system
reduces to a resultant computation as illustrated in the sequel.
The {\it Sylvester resultant method} is the technique most
frequently utilized for determining a common root of two polynomial
equations in one variable. 
However, using the Sylvester method successively to solve a system of multivariate
polynomials proves to be inefficient. 

It is more efficient to eliminate $n$ variables together from $n+1$
polynomials, thus, leading to the notion of the {\it multivariate
resultant.} The three most commonly used multivariate resultant
matrix formulations are those named after {\it Sylvester or Macaulay}
\cite{16.Ca90:1990, 16.CKL89, 16.Mac16:1916},
those named after {\it B\'{e}zout or Dixon}
\cite{16.BuElMo01, 16.Di08:1908Quantics, 16.KpSa95:1995Resultant},
or the {\it hybrid formulation}
\cite{16.DicEmi03, 16.Jouanolou, 16.Khet03}.  
Extending the Sylvester-Macaulay type, we shall emphasize also
{\it sparse resultant} formulations
\cite{16.CE00:2000, 16.GeKaZe94:1994Resultants, 16.Stu91:June1991Elimination}.
For a unified treatment, see \cite{16.EmiMou99jsc}.

The theory of Gr\"{o}bner bases provides powerful tools for performing
computations in multivariate polynomial rings. Formulating the
problem of solving systems of polynomial equations in terms of
polynomial ideals, we will see that a Gr\"obner basis can be
computed from the input polynomial set, thus, allowing for a
form of back substitution in order to compute the common roots.

Although not discussed, it should be noted that the {\it characteristic
set algorithm} can be utilized for solving polynomial systems.
Although introduced for studying algebraic differential
equations~\cite{16.Ri50:1950Differential},
the method was converted to ordinary polynomial rings when developing
an effective method for automatic theorem proving~\cite{16.Wu84:1984SystMath}.
Given a polynomial system $P$, the characteristic set
algorithm computes a new system in triangular form, such that the
set of common roots of $P$ is equivalent to the set of roots of
the triangular system~\cite{16.KpLa92:1992Elimination}.
{\em Triangular systems} have $k_1$ polynomials in a specific
variable, $k_2$ polynomials in this and one more variable,
$k_3$ polynomials in these two and one more variable, and so on, for a
total number of $k_1+\cdots+k_n$ polynomials.


\subsection{Resultant of Univariate Systems}

The question of whether two polynomials $f(x),~g(x)\in \RR{Q}[x]$,
\begin{eqnarray*}
   f(x) & = & f_nx^n + f_{n-1}x^{n-1} + \cdots + f_1x + f_0~,   \\
   g(x) & = & g_mx^m + g_{m-1}x^{m-1} + \cdots + g_1x + g_0~,
\end{eqnarray*}
have a common root leads to a condition that has to be satisfied by
the coefficients of $f, g$. Using a derivation of this
condition due to Euler, the {\it Sylvester matrix} of $f,g$
(which is of dimension $m+n$) can be formulated.
The vanishing of the determinant of the Sylvester matrix, known as
the {\it Sylvester resultant,} is a necessary and sufficient condition
for $f,g$ to have common roots in the algebraic closure of the
coefficient ring. 

As a running example let us consider the following bivariate system 
\cite{16.Laz81:1981}:
\begin{eqnarray*}
   f & = & x^2 + xy + 2x \quad \quad \, + y - 1  = 0~,  \\
   g & = & x^2 \quad \quad + 3x - y^2 + 2y - 1  = 0~.
\end{eqnarray*}
Without loss of generality, the roots of the Sylvester resultant of $f$ and
$g$ treated as polynomials in $y$, whose coefficients
are polynomials in $x$, are the $x$-coordinates of the
common roots of $f,g$. More specifically, the
Sylvester resultant with respect to $y$ is given by the following determinant:
$$
   \det \left[ \begin {array}{ccc}
                      x + 1 & {x}^{2} + 2\, x - 1 & 0            \\
    \noalign{\medskip}0     & x +1                &{x}^{2}+2\,x-1\\
    \noalign{\medskip}-1    &2                    &{x}^{2}+3\,x-1
           \end {array} \right] = -{x}^{3}-2\,{x}^{2}+3\,x~.
$$
An alternative matrix of order $\max\{m, n\}$, named after B\'{e}zout, yields
the same determinant.

The roots of the Sylvester determinant are
$\{-3,0,1\}$. For each $x$ value, one can substitute the
$x$ value back into the original polynomials yielding the
solutions $(-3,1), (0,1), (1, -1)$.
More practically, one can use the Sylvester matrix to reduce system solving
to the computation of eigenvalues and eigenvectors as explained in
``Polynomial System Solving by Using Resultants".


The Sylvester formulations has led to a {\it subresultant theory},
which produced an efficient algorithm for computing
the GCD of univariate polynomials and their resultant, while
controlling intermediate expression swell
\cite{16.Reischert:subresultant:97,16.LickteigRoy:FastCauchy:01}.
Subresultant theory has been generalized to several variables, e.g.\
\cite{16.BusDAn:2004subres, 16.DAnKriSza06}.

\subsection{Resultants of Multivariate Systems}

The solvability of a set of nonlinear multivariate polynomials
is determined by the vanishing of a
generalization of the resultant of two univariate polynomials.
We examine two generalizations: the classical and the
sparse resultants.
Both generalize the determinant of $n+1$ {\em linear}
polynomials in $n$ variables.

The {\it classical resultant} of a system of $n+1$ polynomials with
symbolic coefficients
in $n$ variables vanishes exactly when there exists a common solution in the
{\it projective} space over the algebraic closure of the coefficient
ring~\cite{16.CLO}.
The {\it sparse (or toric) resultant} characterizes solvability of the same
overconstrained system over a smaller
space, which coincides with affine space under certain genericity
conditions~\cite{16.CLO2, 16.GeKaZe94:1994Resultants, 16.Stu91:June1991Elimination}.
The main algorithmic question is to construct a matrix whose
determinant is the resultant or a nontrivial multiple of it.

Cayley, and later Dixon, generalized B\'{e}zout{}'s method 
to a set
$$
  \left\{ p_1 \left( x_1,\ldots,x_n \right), \ldots, p_{n+1}
       \left( x_1,\ldots,x_n \right) \right\}
$$
of $n+1$ polynomials in $n$ variables.
The vanishing of the determinant of the B\'{e}zout--Dixon matrix
is a necessary and sufficient condition for the polynomials
to have a nontrivial projective common root, and also a necessary condition
for the existence of an affine common root
\cite{16.BuElMo01, 16.Di08:1908Quantics, 16.EmiMou99jsc, 16.KpSa95:1995Resultant}.
A nontrivial resultant multiple, known as the
{\it projection operator,} can be extracted via a method
discussed in~\cite[thm. 3.3.4]{16.CaMo95}.
This article, along with~\cite{16.ElMo96}, explain the correlation
between residue theory and the B\'{e}zout--Dixon matrix; the former
leads to an alternative approach for studying and approximating
all common solutions.

Macaulay~\cite{16.Mac16:1916} constructed a matrix whose
determinant is a multiple of the classical resultant; he stated his approach
for a well-constrained system of $n$ homogeneous polynomials in $n$ variables.
The Macaulay matrix simultaneously generalizes
the Sylvester matrix and the coefficient matrix of a system of linear
equations. Like the Dixon
formulation, the Macaulay determinant is a multiple of the resultant.
Macaulay, however, proved that a certain minor of his matrix divides
the matrix determinant to yield the exact resultant
in the case of generic coefficients.  To address arbitrary coefficients,
Canny~\cite{16.Ca90:1990} proposed a general method
that perturbs any polynomial system and extracts a nontrivial
projection operator from Macaulay's construction.

By exploiting the structure of polynomial systems by means of sparse
elimination theory,
a matrix formula for computing the sparse resultant of $n+1$ polynomials in
$n$ variables was given in~\cite{16.CE00:2000}
and consequently improved in~\cite{16.CaPe93, 16.EmCa95}.
Like the Macaulay and Dixon matrices,
the determinant of the sparse resultant matrix, also known as Newton matrix,
only yields a projection operation.
However, in certain cases of bivariate and multihomogeneous
systems, determinantal formulae for the sparse resultant have been 
derived \cite{16.DicEmi03, 16.EmiMan09, 16.Khet03}.
To address degeneracy issues, Canny's perturbation has been
extended in the sparse context~\cite{16.DAndEmir:2001pert}.
D'Andrea~\cite{16.DAnd02} extended Macaulay's rational formula for
the resultant to the sparse setting, thus defining the sparse resultant
as the quotient of two determinants; see \cite{16.EmiKonJsc} for a simplified
algorithm in certain cases.

Here, sparsity means that only certain monomials in each
of the $n+1$ polynomials have nonzero coefficients.
Sparsity is measured in geometric terms, namely, by the {\bf Newton polytope}
of the polynomial, which is the convex hull of the exponent vectors
corresponding to nonzero coefficients.
The {\bf mixed volume} of the Newton polytopes of $n$ polynomials in
$n$ variables is defined as an integer-valued function that bounds
the number of toric common roots of these polynomials~\cite{16.Be75}.
This remarkable bound is the cornerstone of sparse elimination theory.
The mixed volume bound is significantly smaller than the classical
B\'{e}zout{} bound for polynomials with small Newton polytopes but they
coincide for polynomials whose Newton polytope is the unit simplex
multiplied by the polynomial's total degree.
Since these bounds also determine the degree of the sparse and
classical resultants, respectively, the latter has larger degree
for sparse polynomials.
Last, but not least, the classical resultant can identically vanish
over sparse systems, whereas the sparse resultant can
still yield the desired information about their common roots~\cite{16.CLO2}. 

\subsection{Polynomial System Solving by Using Resultants}

Suppose we are asked to find the common roots of a set of
$n$ polynomials in $n$ variables $\{p_1(x_1,\ldots,x_n)$,
$\ldots$, $p_n(x_1,\ldots,x_n)\}$.
By augmenting this set by a generic linear
polynomial~\cite{16.Ca90:1990, 16.CLO2},
we construct the {\it u-resultant} of a given system of polynomials.
The u-resultant is named after the indeterminates $u$,
traditionally used to represent the generic coefficients of the
additional linear polynomial.
The u-resultant factors into linear factors over the
complex numbers, providing the common roots of the given
polynomials equations. The method relies on
the properties of the multivariate resultant, and hence, can
be constructed using either Macaulay's, Dixon's, or sparse formulations.
An alternative approach is to {\it hide} a variable in the
coefficient field~\cite{16.Emi96c, 16.EmiMou99jsc, 16.Ma92}.

Consider the previous example augmented by a generic linear form:
\begin{eqnarray*}
  p_1 & = & x^2 + xy +2x \quad + y - 1 = 0~,                    \\
  p_2 & = & x^2 \quad  + 3x - y^2 + 2y - 1 = 0~,                \\
  p_l & = &  \quad \quad \quad ux \quad \quad + vy + w = 0~.
\end{eqnarray*}

As described in~\cite{16.CKL89},
the following (transposed) Macaulay matrix $M$ corresponds to the u-resultant
of the above system of polynomials:
$$
 M=\left[
\begin{array}{rrrrrrrrrr}
 1&0&0&1&0&0&0&0&0&0\\
 1&1&0&0&1&0&u&0&0&0\\
 2&0&1&3&0&1&0&u&0&0\\
 0&1&0&-1&0&0&v&0&0&0\\
 1&2&1&2&3&0&w&v&u&0\\
 -1&0&2&-1&0&3&0&w&0&u\\
 0&0&0&0&-1&0&0&0&0&0\\
 0&1&0&0&2&-1&0&0&v&0\\
 0&-1&1&0&-1&2&0&0&w&v\\
 0&0&-1&0&0&-1&0&0&0&w\\
\end{array}
\right]~.
$$
It should be noted that
$$
     \det(M) = (u-v+w) (-3u+v+w) (v+w)(u-v)
$$
corresponds to the affine solutions $(1,-1)$, $(-3,1)$, $(0,1)$,
whereas one solution at infinity corresponds to the last factor.

Resultant matrices can also reduce polynomial system solving to
a regular or generalized eigenproblem
(cf.\ ``Matrix Eigenvalues and Singular Values Problems''),
thus, transforming the nonlinear question to a problem in linear algebra.
This is a classical technique that enables us to numerically
approximate all solutions \cite{16.AuSt88,
                     16.CaMa91:1991Resultant, 16.CaMo95,
                     16.Emi96c, 16.EmiMou99jsc}.
For demonstration, consider the previous system and its resultant matrix
$M$.
The matrix rows are indexed by the following row vector of monomials in the
eliminated variables:
$$
  {\bf v} = \left[ x^3, x^2 y, x^2, x y^2, x y, x, y^3, y^2, y, 1 \right]~.
$$
Vector ${\bf v}M$ expresses the polynomials indexing the columns of $M$,
which are multiples of the three input polynomials by various monomials.
Let us specialize variables $u$ and $v$ to random values. Then
$M$ contains a single variable $w$ and is denoted $M(w)$.
Solving the linear system
${\bf v}M(w) ={\bf 0}$ in vector ${\bf v}$ and in scalar
$w$ is a generalized eigenproblem,
since $M(w)$ can be represented as $M_0 + w M_1$, where $M_0$ and $M_1$
have
numeric entries.
If, moreover, $M_1$ is invertible, we arrive at the following
eigenproblem:
$$
 {\bf v} \left( M_0 + w M_1 \right) = {\bf 0}
   \Longleftrightarrow
     {\bf v} \left( - M_1^{-1} M_0 - w I \right) ={\bf 0}
       \Longleftrightarrow
          {\bf v} \left( - M_1^{-1} M_0 \right) =  w {\bf v}~.
$$
For every solution $(a,b)$ of the original system, there is
a vector ${\bf v}$
among the computed eigenvectors, which we evaluate at $x=a,~y=b$
and from which the solution can be recovered by division~\cite{16.Emi96c}.
As for the eigenvalues, they correspond to the values of $w$ at the
solutions; see~\cite{16.Emir01dags} on numerical issues, and an implementation.

An alternative method for approximating or isolating all real roots
of the system is to use the so-called
Rational Univariate Representation (RUR) of algebraic numbers
\cite{16.Can88pspace, 16.Rouillier99}.
This allows us to express each root coordinate as the value of a 
univariate polynomial, evaluated over an
algebraic number, which is specified as a
solution of a single polynomial equation.
All polynomials involved in this approach are derived from the resultant.

The resultant matrices are sparse and have quasi Toeplitz/Hankel structure
(also called multilevel Toeplitz/Hankel structure), which enables their
fast multiplication by vectors. By combining the latter property
with various advanced nontrivial methods of multivariate polynomial root-finding,
substantial acceleration of the construction and computation  of the
resultant matrices and approximation of the system's solutions was achieved
in~\cite{16.BoMoPa98/00, 16.EmPa02, 16.EmPa03/05, 16.MoPa98, 16.MoPa00, 16.MoPaRu02/03}.

A comparison of the resultant formulations can be found, e.g.,
in~\cite{16.EmiMou99jsc, 16.KpLa92:1992Elimination, 16.Ma92}.
The multivariate resultant formulations have been
used for diverse applications such as
{\it algebraic and geometric reasoning}~\cite{16.CaMo95, 16.DickStur02, 16.Ma92},
including separation bounds for the isolated roots of arbitrary polynomial
systems~\cite{16.EMT},
{\it robot kinematics}~\cite{16.DanEmi01icra, 16.RaRo95, 16.Ma92},
and {\it nonlinear computational geometry,
computer-aided geometric design and, in particular, implicitization}
\cite{16.BusDAn:2004subres, 
16.ChCoLi05, 16.EmiKalKonBa, 16.EmiTzo08, 16.HSW1997}. 

\subsection{Gr\"{o}bner Bases}

Solving systems of nonlinear equations can be formulated in terms of polynomial
ideals~\cite{16.CLO, 16.GruPfi02, 16.KreRob00}. 
The {\it ideal} generated by a system of polynomials
$p_1,\ldots,p_r$ over $\RR{Q}[x_1,\ldots,x_n]$ is the set of all linear
combinations
$$
  \left( p_1,\ldots,p_r \right) = \left\{ h_1p_1 + \cdots + h_rp_r
     \mid h_1,\ldots,h_r \in \RR{Q} \left[ x_1,\ldots,x_n \right] \right\}~.
$$
The algebraic variety of $p_1,\ldots,p_r\in \RR{Q}[x_1, \ldots,x_n]$
is the set of their common roots,
$$
  V \left( p_1,\ldots,p_r \right) =
    \left\{ \left( a_1,\ldots,a_n \right) \in \RR{C}^n \mid p_1
      \left( a_1, \ldots,a_n \right) = \ldots =
        p_r \left( a_1,\ldots,a_n \right) = 0 \right\}~.
$$
A version of the {\it Hilbert Nullstellensatz} states that
$$
   V \left( p_1,\ldots,p_r \right) = \mbox{the empty set }
     \emptyset
       \Longleftrightarrow 1 \in \left( p_1,\ldots,p_r \right) \;
         {\rm over}\;
          \RR{Q} \left[ x_1,\ldots,x_n \right]~,
$$
which relates the solvability of polynomial systems to the
ideal membership problem.

A term $t = x_1^{e_1}x_2^{e_2}\ldots x_n^{e_n}$ of a polynomial
is 
a 
product of powers with ${\rm deg}(t)=e_1+\cdots+e_n$.
In order to add needed structure to the polynomial ring we
will require that the terms in a polynomial be ordered in an
admissible fashion~\cite{16.CLO, 16.GeCzLa92:1992Algebra}.
Two of the most common admissible orderings are the
{\bf lexicographic order}
(${\prec}_l$), where terms are ordered as in a dictionary, and the
{\bf degree order}
(${\prec}_d$), where terms are first compared by their degrees with equal
degree terms compared lexicographically. A variation to the lexicographic
order is the {\it reverse lexicographic order,} where the lexicographic
order is reversed.

Much like a polynomial remainder process,
the process of polynomial reduction involves subtracting a multiple of
one polynomial from another to obtain a smaller degree
result~\cite{16.CLO, 16.GruPfi02, 16.KreRob00}. 
A polynomial $g$ is said to be reducible with respect to a set
$P = \{p_1,\ldots,p_r\}$ of polynomials if it can be reduced by
one or more polynomials in $P$. When $g$ is no longer reducible by
the polynomials in $P$, we say that $g$ is {\it reduced} or is
{\it a normal form} with respect to $P$.

For an arbitrary set of basis polynomials, it is possible that
different reduction sequences applied to a given polynomial $g$
could reduce to different normal forms. A basis
$G\subseteq \RR{Q}[x_1,\ldots,x_n]$ is a {\it Gr\"{o}bner basis} if and
only if every polynomial in $\RR{Q}[x_1,\ldots,x_n]$ has a unique normal
form with respect to $G$. Buchberger~\cite{16.Bu65:Fall1965thesis,
16.Bu76:1976SIGSAM, 16.Bu85:1985groebnersurvey}
showed that every basis for an
ideal $(p_1,\ldots,p_r)$ in $\RR{Q}[x_1,\ldots,x_n]$ can be converted
into a Gr\"obner basis
$\{p_1^*,\ldots,p_s^*\} = GB (p_1,\ldots,p_r)$,
concomitantly designing an algorithm that transforms an arbitrary ideal
basis into a Gr\"{o}bner basis.
Another characteristic of Gr\"{o}bner bases is that by using the above
mentioned reduction process we have
$$
    g\in \left( p_1\ldots,p_r \right)
\Longleftrightarrow g \; {\rm mod} \; \left( p_1^*,\ldots,p_s^* \right) = 0~.
$$
Further, by using the Nullstellensatz it can be shown that
$p_1\ldots,p_r$ viewed as a system of algebraic equations is
solvable if and only if $1 \not\in GB(p_1,\ldots,p_r)$.

Depending on which admissible term ordering is used in the
Gr\"{o}bner bases construction, an ideal can have different
Gr\"{o}bner bases. However, an ideal cannot have different
(reduced) Gr\"{o}bner bases for the same term ordering.
Any system of polynomial equations can be solved using a
lexicographic Gr\"{o}bner basis for the ideal generated by the
given polynomials. It has been observed, however, that Gr\"{o}bner
bases, more specifically lexicographic Gr\"{o}bner bases, are hard
to compute~\cite{16.MayMey81}.
In the case of zero-dimensional ideals, those whose varieties have only
isolated points, a change of basis algorithm was outlined
in~\cite{16.FGLM89:1993faugere}, which can be utilized for solving:
one computes a Gr\"{o}bner basis for the ideal generated by a system of
polynomials under a degree ordering. The so-called {\it change
of basis algorithm} can then be applied to the degree ordered Gr\"{o}bner
basis to obtain a Gr\"{o}bner basis under a lexicographic ordering.
Significant progress has been achieved in the algorithmic
realm by Faug\`{e}re~\cite{16.Faug99, 16.Faug02}.

Another way to finding all common real roots is by means of RUR;
see the previous section.
All polynomials involved in this approach can be derived from the 
Gr\"{o}bner basis.
A rather recent development concerns the generalization of
Gr\"{o}bner bases to {\em border bases}, which contain all information
required for system solving but can be computed faster and seem to be
numerically more stable \cite{16.KreRob00, 16.MouTre02, 16.stetter04:numer,16.MouTre08}.

Turning to Lazard's example in form of a polynomial basis,
$$
\begin{array}{ccllllll}
p_1  &=   &x^2   &+ xy   &+ 2x   &      &+ y   &- 1~,             \\
p_2  &=   &x^2   &       &+ 3x   &- y^2 &+ 2y  &- 1~,
\end{array}
$$
one obtains (under lexicographical ordering with $x {\prec}_l y$) a
Gr\"{o}bner basis in which the variables are triangulated such that the
finitely many solutions can be computed via back substitution:
$$
\begin{array}{ccllllll}
{p_1}^*   &=    &x^2  &   &+ 3x   &    &+ 2y  &- 2~,              \\
{p_2}^*   &=    &     &xy &- \, x &    &- y   &+ 1~,              \\
{p_3}^*   &=    &     &   &       &y^2 &      &- 1~.
\end{array}
$$
The final univariate polynomial
has minimal degree, whereas the polynomials used in the back
substitution have total degree no larger than the number of roots.
As an example,
$x^2y^2$ is reduced with respect to the previously
computed Gr\"{o}bner basis $\{p_1^*,p_2^*,p_3^*\}=GB(p_1,p_2)$ along
two distinct reduction paths, both yielding $-3x-2y+2$ as the normal form.

There is a strong connection between lexicographic Gr\"{o}bner bases and the
previously mentioned resultant techniques. For some types of input
polynomials, the computation of a reduced system via resultants might be
much faster than the computation of a lexicographic Gr\"{o}bner basis.

Gr\"{o}bner bases can be used for many polynomial ideal theoretic operations
\cite{16.Bu85:1985groebnersurvey, 16.Cox07}.
Other applications include 
computer-aided geometric design~\cite{16.HSW1997},
polynomial interpolation~\cite{16.LaSa95:1995},
coding and cryptography~\cite{16.FaLePe08}, 
and robotics~\cite{16.FaLa95}.

\section{Research Issues and Summary}\label{16.sec5}

Algebraic algorithms deal with numbers, vectors,  
matrices, polynomials, formal power series,
exponential and differential polynomials, rational functions,
algebraic sets, curves and surfaces.
In this vast area, manipulations with matrices and polynomials, 
in particular the solution of a polynomial equation and 
linear and polynomial systems of equations,  
are most fundamental in modern computations in Sciences, Engineering,
and Signal and Image Processing.
We reviewed the state of the art for the solution
of these three tasks and gave pointers
to the extensive bibliography.

Among numerous interesting and important research directions of the topics 
in Sections~2 and~3,
we wish to cite computations with structured matrices, 
including their applications to polynomial root-finding, currently of growing
interest, and new techniques for randomized preprocessing 
for matrix computations, evaluation of resultants
and polynomial root-finding. 

Section~\ref{16.sec:nonlin} of this chapter has briefly
reviewed polynomial system solving based on
resultant matrices as well as Gr\"{o}bner bases.
Both approaches are currently active. This includes practical applications
to small and medium-size systems.
Efficient implementations that handle the nongeneric cases,
including multiple roots and nonisolated solutions,
is probably the most crucial issue today in relation to resultants.
The latter are also studied in relation to a more general object,
namely the discriminant of a well-constrained system, which characterizes
the existence of multiple roots.
Another interesting current direction is algorithmic improvement by
exploiting the structure of the polynomial systems, including sparsity,
or the structure of the encountered matrices, for both resultants and Gr\"{o}bner bases.

\section{Defining Terms}

\begin{description}

\item[Characteristic polynomial:]
Shift an input matrix $A$ by subtracting  the identity matrix $xI$
scaled by variable $x$.
The determinant of the resulting matrix is the characteristic polynomial
of the matrix $A$. Its roots coincide with the eigenvalues of the shifted
matrix $A-xI$.

\item[Condition number of a matrix] is a scalar $\kappa$ which
grows large as
  the matrix approaches a singular matrix;
then numeric inversion becomes an
ill-conditioned problem. 
 $\kappa$ \texttt{OUTPUT ERROR NORM} $\approx $ \texttt{INPUT ERROR NORM}.

\item[Degree order:] An order on the terms in a multivariate polynomial;
for two variables $x$ and $y$ with $x \prec y$ the ascending chain of
terms is $1$ $\prec$ $x$ $\prec$ $y$ $\prec$ $x^2$ $\prec$ $xy$ $\prec$
$y^2$ $\cdots$.

\item[Determinant:] A polynomial in the entries of a square matrix whose
 value is invariant in adding to a row (resp. column) any linear combination 
of other rows (resp. columns). $\det (AB)=\det A\cdot \det B$ for
a pair of square matrices $A$ and $B$, $\det B=-\det A$
if the matrix $B$ is obtained by interchanging
a pair of adjacent rows or columns of a matrix $A$, $\det A\neq 0$
if and only if a matrix $A$ is invertible. Determinant of a block diagonal 
or block triangular matrix is the product
of the diagonal blocks, and so $\det A=(\det A_{0.0})\det S$
under~(\ref{eqbl}). One can compute a determinant by using  
these properties and matrix factorizations,
e.g., recursive factorization~(\ref{eqbl}).

\item[Gr\"obner basis:] Given a term ordering, the Gr\"obner basis
of a polynomial ideal is a generating set of this ideal, such that the
(multivariate) division of any polynomial by the basis has a unique remainder.

\item[Lexicographic order:] An order on the terms in a multivariate
polynomial; for two variables $x$ and $y$ with $x \prec y$ the
ascending chain of terms is $1$ $\prec$ $x$ $\prec$ $x^2$ $\prec$
$\cdots$ $\prec$ $y$ $\prec$ $xy$ $\prec$ $x^2 y$ $\cdots$ $\prec$
$y^2$ $\prec$ $xy^2$ $\cdots$.

\item[Matrix eigenvector:] A column vector ${\bf v}$ such that
$A {\bf v} = \lambda {\bf v}$, for a
square matrix $A$ and the associated eigenvalue $\lambda$.  
A generalized eigenvector ${\bf v}$ satisfies
the equation $A{\bf v} = \lambda B{\bf v}$ for two
square matrices
$A$ and $B$ and the associated eigenvalue $\lambda$.
Both definitions extend to row vectors that premultiply the
associated  matrices.

\item[Mixed volume:] An integer-valued function of $n$ convex polytopes in
$n$-di\-men\-sion\-al Euclidean space.
Under proper scaling, this function bounds the number of toric complex roots
of a well-constrained polynomial system, where the convex polytopes are
defined to be the Newton polytopes of the given polynomials.

\item[Newton polytope:] 
The convex hull of the exponent vectors corresponding to terms with
nonzero coefficients in a given multivariate polynomial.

\item[Ops:] Arithmetic operations, i.e., additions, subtractions,
multiplications,
or divisions; as in {\bf flops,} i.e., floating point operations.

\item[Resultant:] A polynomial in the coefficients of a system of $n$ polynomials with
  $n+1$ variables, whose vanishing is the minimal
 necessary and sufficient condition for the existence of a solution of the system.

\item[Separation bound:] The minimum distance between two (complex) roots of
a univariate polynomial.

\item[Singularity:] A square matrix is singular if its product
with some nonzero
matrix  
is the zero matrix. Singular matrices do not have inverses.

\item[Sparse matrix:] A matrix whose zero entries are much more numerous
than its  nonzero entries.

\item[Structured matrix:] A matrix whose every entry can be derived by
a formula depending on a smaller number of parameters, typically on 
$O(m+n)$ parameters for an $m\times n$ matrix, 
as opposed to its $mn$ entries. 
For instance, an $m\times n$ Cauchy matrix
has $1\over {s_i-t_j}$ as the entry in row $i$ and column $j$
and is defined by
$m+n$ parameters $s_i$  and $t_j$, $i=1,\dots,m$; $j=1,\dots,n$.
Typically a structured matrix can be multiplied by a vector in nearly linear 
arithmetic time.

\end{description}



\section*{Further Information}

The books and special issues of journals~\cite{16.AHU74:1974,
                16.BaSh96,
                16.BiPa94:1994algorithms,
                16.BoMu75:1975,
                16.BuClSh97,
		16.DicEmi05book,
		16.EmMoPa04,
                16.GeCzLa92:1992Algebra,
		16.Pan01,
		16.stetter04:numer,
                16.Zi93:1993}
provide a broader introduction to the general subject
and further bibliography.

There are well-known
libraries and packages of subroutines for the most popular numerical matrix
computations,
in particular, \cite{16.Do78:1978LAPACK} for solving linear systems of
equations,~\cite{16.Ga72:1972EISPACK},  \cite{16.Sm70:1970EISPACK}, 
ARPACK, and PARPACK for
approximating matrix eigenvalues, and~\cite{16.An92:1992LAPACK} for both
of the two latter computational problems.
Comprehensive treatment of numerical matrix
computations and extensive
bibliography
can be found in~\cite{16.GoLo96, 16.Ste98}, and there are
many more specialized books on
them~\cite{16.BDDR00,
	   16.BBC93, 
	   16.DDSV98,
	   16.GL81,	
	   16.Green97,
           16.High02,
           16.Pa80:1980eigenvalue,
	   16.TB97,
           16.Wi65:1965eigenvalue}
as well as many survey
articles~\cite{16.HePe91:1991sparse,
               16.OrVo85:1985parallel,
               16.Pa92}
and thousands of research articles.
Further applications to the graph and combinatorial computations
related to linear algebra are cited in
``Some Computations Related to Matrix Multiplication''
and \cite{16.Pa93}.


On parallel matrix computations see 
\cite{16.GiSc92, 16.GoLo96, 16.KaPa91:1991spaa,
16.KaPa92:1992focs, 16.PaPr95:1995workspeedmatrix}
assuming general input matrices, 
\cite{ 
16.GiSc92, 16.HePe91:1991sparse, 16.Pa93, 16.PaRe93}
assuming sparse inputs, \cite{16.DDSV98} assuming 
banded inputs, and \cite{16.BiPa94:1994algorithms, 16.Pan96corr, 16.Pan01}
assuming dense structured inputs.
On Symbolic-Numeric algorithms, see the books
\cite{16.BiPa94:1994algorithms, 16.Pan01, 16.WaZh07},
surveys \cite{16.Pa92, 16.Pa97:1997survey, 16.Pan98},
special issues~\cite{16.EmMoPa04, 16.BiPaVe08, 16.KMP11, 16.KMPZ}, and the bibliography therein.
For the general area of exact computation and the theory behind algebraic
algorithms and computer algebra, see
\cite{16.BPR03, 16.bcla-casac-83, 16.CLO, 16.CLO2,
16.DicEmi05book, 16.GaGe03, 16.GeCzLa92:1992Algebra, 
16.MignotteStefanecu,
16.Winkler96, 16.Yap2000, 16.m-mca-92, 16.Zi93:1993}.

There is a lot of generic software packages for exact computation,
\textsc{synaps} 
\cite{16.mptt-mega-2005}, a \texttt{C++} open source library devoted to
symbolic and numeric computations with polynomials, algebraic numbers
and polynomial systems, which has been evolving
into the \textsc{realroot} package of the open source
computer algebra system \textsc{mathemagix}; 
\textsc{ntl} 
a high-performance
\texttt{C++} library providing data structures and algorithms
for vectors, matrices, and polynomials over the integers and finite fields,
and \textsc{exacus} 
\cite{16.exacus:05}, a \texttt{C++} library for curves and surfaces that
provides exact methods for solving polynomial equations.  
A highly efficient tool is \textsc{FGb} 
for Gr{\"o}bner basis, and \textsc{RS} for the rational
univariate representation, and real solutions of
systems of polynomial equations and inequalities.  Finally, 
\textsc{LinBox} 
\cite{16.linbox} is a \texttt{C++} library that provides exact
high-performance implementations of linear algebra algorithms.

This chapter does not cover the area of polynomial factorization. We refer 
the interested reader to 
\cite{16.GaGe03, 16.LLL82:1982LLLlovasz, 16.NV10},
and the bibliography therein.

The {\it SIAM Journal on Matrix Analysis and Applications} and
{\it Linear Algebra and Its Applications} are specialized on Matrix 
Computations,
{\it Mathematics of Computation} and {\it Numerische Mathematik} are leading 
among numerous other good journals on numerical computing.

The {\it Journal of Symbolic Computation} and the
{\it Foundations of Computational Mathematics}
specialize on topics in Computer Algebra,
which are also covered in the {\em Journal of Computational Complexity},
the {\em Journal of Pure and Applied Algebra}
and, less regularly, in the {\em Journal of Complexity}. 
{\it Mathematics for Computer Science}
and {\it Applicable Algebra in Engineering, Communication and Computing}
are currently dedicated to the subject of the chapter as well.
{\em Theoretical Computer Science} has
become more open to algebraic--numerical and algebraic--geometric subjects
\cite{16.BiPaVe08, 16.BuElMoa, 16.EmMoPa04, 16.KMP11}.

The annual {\it International Symposium on Symbolic and Algebraic
Computation (ISSAC)} is the main conference in computer algebra;
these topics are also presented at the bi-annual Conference {\em MEGA}
and the newly founded SIAM conference on Applications of Algebraic Geometry.
They also appear, 
in the annual {\it ACM Conference on Computational Geometry}, as well as
at various Computer Science conferences, including SODA, FOCS, and STOC.

Among many conferences on numerical computing, most comprehensive ones
are organized under the auspices of SIAM and ICIAM.
The International Workshop on Symbolic-Numeric Algorithms can be traced
back to 1997 (SNAP in INRIA, Sophia Antipolis) and a
special session in IMACS/ACA'98 Conference 
in Prague, Czech Republic, in 1998 \cite{16.Pan98}.
It restarted in Xi'an, China,
2005; Timishiora, Romania,  2006 (supported by IEEE), and 
London, Ontario, Canada, 2007 (supported by ACM).  
The topics of Symbolic-Numerical Computation are also represented at the
conferences on the {\em Foundations of Computational Mathematics (FoCM)}
(meets every 3 years) and quite often at ISSAC.
\end{document}